\documentclass[superscriptaddress,amsmath,twocolumn,amssymb,aps,showpacs,prl]{revtex4-1}
\usepackage[utf8]{inputenc}
\usepackage{graphicx}
\usepackage{bm}
\usepackage{amssymb}
\usepackage{amsfonts}
\usepackage{color}
\usepackage{natbib}
\usepackage[caption=false]{subfig}
\newcommand{\vep}{\varepsilon}

\newcommand{\bqa}{\begin{eqnarray}}
\newcommand{\eqa}{\end{eqnarray}}
\newcommand{\nn}{\nonumber}

\usepackage{atbegshi,picture}
\usepackage{lipsum}

\AtBeginShipoutNext{\AtBeginShipoutUpperLeft{%
		\put(\dimexpr\paperwidth-6.cm\relax,-0.5cm){\makebox[0pt][r]{{Submitted to Special Issue Localisation 2020 of Annals of Physics}}}%
	}}

\begin{document}

\date{\today}

\title{Inhomogeneous Kondo destruction by RKKY correlations}

\author{Kyung-Yong Park}
\affiliation{Department of Physics, POSTECH, Pohang, Gyeongbuk 37673, Korea}

\author{Iksu Jang}
\affiliation{Department of Physics, POSTECH, Pohang, Gyeongbuk 37673, Korea}

\author{Ki-Seok Kim}
\affiliation{Department of Physics, POSTECH, Pohang, Gyeongbuk 37673, Korea}
\affiliation{Asia Pacific Center for Theoretical Physics (APCTP), Pohang, Gyeongbuk 37673, Korea}

\author{S. Kettemann}
\affiliation{Division of Advanced Materials Science, POSTECH, Pohang, Gyeongbuk 37673, Korea}
\affiliation{Department of Physics and Geoscience, Jacobs University
  Bremen, Bremen 28759, Germany}

%
%

\begin{abstract}
The competition between  the indirect exchange interaction (IEC) of  magnetic impurities in metals and  the Kondo effect gives rise to a rich quantum phase diagram, the Doniach Diagram \cite{doniach}. A Kondo screened phase is separated from a spin ordered phase when the local exchange coupling $J$ and the concentration of magnetic moments $n_{M}$ are varied. In disordered metals, both the Kondo temperature and the IEC are widely distributed due to the scattering of the conduction electrons from the impurity potential. Therefore, it is a question of fundamental importance, how this Doniach diagram is modified by the disorder, and if one can still identify separate phases.
Recently,
Ref. \cite{Nejati2017}   investigated the effect of Ruderman-Kittel-Kasuya-Yosida (RKKY) correlations on the Kondo effect of two magnetic impurities, renormalizing the Kondo interaction based on the Bethe-Salpeter equation and performing the poor men's renormalization group (RG) analysis with the RKKY-renormalized Kondo coupling. In the present study, we extend this theoretical framework,  allowing for different  Kondo temperatures  of  two RKKY-coupled  magnetic  impurities due to different local exchange couplings and density of states. As a result, we find  that the
smaller one of the two  Kondo temperatures is suppressed more strongly by the RKKY interaction, thereby enhancing their initial inequality.
%
%
In order to find out if  this relevance of   inequalities between   Kondo temperatures modifies the distribution of the Kondo temperature in a  system of a finite density of randomly distributed magnetic impurities, we
present an extension of the RKKY coupled Kondo RG equations. We discuss the implication of these results for the  interplay between Kondo coupling and RKKY interaction in disordered electron systems and the Doniach diagram in disordered electron systems.
\end{abstract}

\maketitle

\section{Introduction}

The interplay of strong correlations and disorder leads to new phenomena and
remains a challenge for condensed matter theory. Magnetic impurities in metals
 stir up the electronic Fermi liquid and  cause
a strong enhancement of the resistivity below the Kondo temperature $T_{K}$.
Impurities  result in Anderson localisation and lead to an
exponential increase of the resistivity at  low electron densities. The
interplay of  the Kondo effect with Anderson localisation has  only recently
received increased attention although  the interplay between spin correlations
    and disorder effects is relevant for many materials, including
   doped semiconductors like Si:P close to the metal-insulator transition \cite{lohneysen},
   and typical heavy Fermion systems like  materials with
     4f or 5f atoms, notably Ce, Yb, or U \cite{miranda}.
       Many of these materials  show a remarkable  magnetic quantum  phase transition
        which can be understood by the competition between
        indirect  exchange interaction, the  Ruderman-Kittel-Kasuya-Yoshida\, (RKKY) interaction
between localised magnetic moments \cite{rkky},
as mediated by the conduction electrons,
 and their Kondo screening.
   Thereby, one finds a suppression of long range magnetic order when the
    exchange coupling $J$ is increased and the Kondo screening wins
    over the RKKY coupling.
      This results  in a typical quantum phase diagram with
       a quantum critical point
       where  the $T_c$  of the magnetic phase
       is vanishing, the Doniach diagram \cite{doniach}.
  In any material there is some degree of disorder.
   In doped semiconductors it arises from the random positioning of the dopants themselves, in the heavy Fermion metals it may arise from structural defects
    or atomic defects.
     As noted  already early \cite{pwanderson}, the physics of
      random systems is fully described only by
       probability distributions, not just averages.
             Thus, for electron  systems with random local magnetic moments
             the derivation of physical properties
             requires the  knowledge  of  distribution functions of
             the Kondo temperature and the RKKY coupling
             \cite{mott,BhattFisher92}.

Electron systems with
     onsite  interaction $U$ and a disorder potential $V$
are modeled by    the Anderson-Hubbard  Hamiltonian,
\begin{eqnarray} \label{dam}
H = \sum_{\langle i,j\rangle, \sigma} t_{i j} c^+_{i \sigma} c_{j \sigma }+ \sum_{i,\sigma} (  V_{i, \sigma} - \mu)
 \hat{n}_{i \sigma} + U \sum_{i }\hat{n}_{i +} \hat{n}_{i -},
\end{eqnarray}
where $\hat{n}_{i \sigma} = c^+_{i \sigma} c_{i \sigma }$
 and $c^+_{i \sigma}, c_{i \sigma }$ are  Fermion creation
  and annihilation operators
  at dopant sites $i$ with spin $\sigma= \pm.$      Onsite energies $V_{i, \sigma}$
   are distributed randomly with
    vanishing   average value
 $\langle V_{i, \sigma} \rangle =0$.   $\mu$ is the chemical potential,
  which is for uncompensated doping at $\mu= U/2$.
  This model has been studied mostly in 2 dimensions,
  with  numerical methods,
 including quantum Monte Carlo \cite{Byczuk2011,Ulmke1997,Pezzoli2010}, dynamical mean field theory based approaches \cite{Byczuk2009,Ulmke1995,Aguiar2006,Aguiar2009,Kotliar2003,Aguiar2013,Byczuk2005}, and Hatree-Fock based approaches \cite{Milovanovic1989,Sachdev1998,Tusch1993},
 and most recently a typical medium dynamics cluster approximation \cite{Jarrell2017,Jarrell2014}.  In that work, the quasiparticle self energy has been
  derived as function of the excitation energy $ \omega,$
     $Im \Sigma ( \omega) \sim \omega^{\alpha_{\Sigma}}$
   and
   found to  have non-Fermi liquid  behavior with power $\alpha_{\Sigma} (W) < 2$,
    which becomes smaller with stronger disorder amplitude $W$.

\section{Doniach Phase Diagram in Disordered Systems}

 When there is a   density of  magnetic impurities
$
  n_{\textrm{imp}} =  R_{}^{-d}
$ with   $R$
the average distance between two magnetic moments,
there is a  critical density $n_{c}$ below which the Kondo effect is
dominant in the competition with RKKY interaction.
When a density is higher than  $n_{c}$ magnetic clusters start to form at some sites.
In an electron  system without disorder the critical density above which the magnetic impurities are
coupled with each  other is found  from the condition that
$|J^{0}_{\textrm{RKKY}} (R_c) | = T_K$.
For example in a 2D sample with
$|J^{0}_{\textrm{RKKY}}|_{k_F R \gg 1} = J^2 \frac{m}{8 \pi^2 k_F^2 R^2}$ and
$T_K = c \vep_F \exp (- D_0/J)$, where $k_F$ is the Fermi momentum and $c \approx 1.14$, the
 critical electron density is found to be
$  n_c = 16 \pi^2 c \frac{\vep_F^2}{J^2}  \exp (-\frac{ D_0}{J})$,
where $2 D_0$ is the electron band  width.
%
 In a disordered system  the Doniach diagram
 is a result of the competition
     between the  Kondo temperature $T_{K i}$
      at a certain site ${\bf r}_i$ and the RKKY coupling
      $J_{\textrm{RKKY}}({\bf r}_{i j})$
      at that
      site
      with other magnetic impurities  located at  sites
      ${\bf r}_j$  at a distance ${\bf r}_{i j}$.
      Thus,   the ratio of
these energy scales\,$J_{\textrm{RKKY}}({\bf r}_{i j}) / T_{K i}$,
is in general widely distributed for a given  disordered
sample so that the full distribution function of these energy  scales is needed to determine the Doniach diagram. In Ref.  \cite{HYLee2014} this problem has been studied, calculating $J_{\textrm{RKKY}}({\bf r}_{i j})$
and $T_{K i}$ each separately in a disordered system as function of the local density of states at sites  ${\bf r}_i$ and  ${\bf r}_j$.


Recently, however, Nejati et al. found
from  renormalization group equations for  a  Kondo  lattice  incorporating selfconsistently the RKKY coupling between magnetic moments \cite{Nejati2017}, that the Kondo temperature is decreased as the RKKY coupling is increased, and that  the Kondo screening is quenched beyond  a critical RKKY coupling.

The effective Kondo coupling $g_i$  of the Kondo impurity at site ${\bf r}_i$ was shown in Ref. \cite{Nejati2017}, to  follow renormalization group equations which are modified by the RKKY coupling as
\bqa && \frac{d g_i}{d \ln D} = - 2 g_i^{2} \Big( 1 - y_i g_{0}^{2} \frac{D_{0}}{T_{K}} \frac{1}{\sqrt{1 + (D/T_{K})^{2}}} \Big) . \eqa
Here, $g_i = \rho(\mu) J_i$ is the dimensionless Kondo coupling constant with the density of states at the chemical potential $ \rho(\mu)$. $D$ is the effective band cutoff  for the renormalization group flow. The first term  in the right hand side is the one-loop $\beta$ function without RKKY interactions. The second term  results from the RKKY correction for the Kondo coupling constant, where $g_{0} =\rho(\mu) J_0 $ is the bare Kondo interaction and $D_{0}$ is the bare bandwidth. $y_i$ is the effective dimensionless RKKY interaction strength at site ${\bf r}_i$, given by \cite{Nejati2017}
\bqa && y_i = - \frac{8 W}{\pi^2 \rho(\mu)^2} {\rm Im} \sum_{j \neq i} e^{i {\bf k}_F {\bf r}_{i j}} G^R_c ({\bf r}_{i j}, \mu) \Pi ({\bf r}_{i j}, \mu) ,
\label{y}
\eqa
%
%
where $W$ is the Wilson ratio as determined by the Bethe Ansatz solution of the Kondo problem \cite{bethe}. $G^R_c ({\bf r}_{i j})$ is the single particle  propagator in the conduction band from site ${\bf r}_{i}$ to ${\bf r}_{j}$.
 The summation is over all other magnetic moments at positions ${\bf r}_{j}$.
 $\Pi ({\bf r}_{i j}, \mu)$ is the RKKY  correlation function of  conduction electrons between sites  ${\bf r}_{i}$ and ${\bf r}_{j}$.
  $y_i$ is found to be always positive \cite{Nejati2017}, while the RKKY correlation function can be positive or negative.

It is interesting to observe that the effective Kondo interaction renormalized by the RKKY interaction is a function of $D/T_{K}$, where $D$ is the renormalization group energy scale and $T_{K}$ is the renormalized Kondo temperature to be determined self-consistently. It turns out that this functional form originates from the spin susceptibility of the magnetic impurity, given by the Bethe Ansatz solution.

For two magnetic moments in a clean system, where the
bare couplings $g_0$ are the same at both sites, and $y_i=y$,
one can solve this differential equation and obtains \cite{Nejati2017}
\bqa && \frac{1}{g} - \frac{1}{g_{0}} = 2 \ln \Big( \frac{D}{D_{0}} \Big) - y g_{0}^{2} \frac{D_{0}}{T_{K}} \ln \Big( \frac{\sqrt{1 + (D/T_{K})^{2}} - 1}{\sqrt{1 + (D/T_{K})^{2}} + 1} \Big) . \nn \eqa
When the energy scale coincides with the Kondo temperature, i.e., $D \rightarrow T_{K}$, the effective Kondo interaction diverges $g \rightarrow \infty$. As a result, one can find a self-consistent equation for the effective Kondo temperature as a function of the RKKY interaction,
\bqa && \frac{T_{K}(y)}{T_{K}(0)} = \exp\Big(- y \alpha g_{0}^{2} \frac{D_{0}}{T_{K}(y)} \Big) , \eqa
where $T_{K}(0) = D_0 \exp (-1/(2g_0))$ is the bare Kondo temperature in the absence of the RKKY interaction and the numerical constant is $\alpha = \ln (\sqrt{2} + 1)$. It turns out that the RKKY interaction gives rise to abrupt destruction for the Kondo effect at the critical
 coupling \cite{Nejati2017}
\begin{equation} \label{yc}
y_c= T^0_{K}/(\alpha~ e g_0^2 D_0 ).
\end{equation}

Here, we extend this theoretical framework,
to allow for inhomogenous  local density of states
    at different sites in a  disordered system and
 thereby  different bare Kondo temperatures,  $T^0_{K i} = D_0 \exp (-1/(2g^0_i))$.

 Let us start by considering
 two magnetic moments at sites ${\bf r}_1$ and ${\bf r}_2$ with exchange coupling
   $J^0_{1}$ and $J^0_{2}$ and  with local  density of states
   $\rho(r_1),\rho(r_2)$, yielding the  bare
   dimensionless local coupling parameters
\begin{align} \label{g0}
g^0_{i}=\rho(r_i) J^0_{i},
\end{align}
for $i=1,2.$
Then, the renormalization group $\beta$ functions are found to be given by
\bqa  \frac{d g_{1}}{d \ln D} &=& - 2 g_{1}^{2} \Big( 1 - y g^0_{1} g^0_{2} \frac{D_{0}}{T_{K 2}} \frac{1}{\sqrt{1 + (D/T_{K 2})^{2}}} \Big) , \nonumber  \\   \frac{d g_{2}}{d \ln D} & = &- 2 g_{2}^{2} \Big( 1 - y g^0_{1} g^0_{2} \frac{D_{0}}{T_{K 1}} \frac{1}{\sqrt{1 + (D/T_{K 1})^{2}}} \Big),
\label{rg2}
\eqa
where $y_1=y_2=y$ is given by
\bqa && y = \frac{8 W}{\pi^2 \rho(\mu)^2} {\rm Im}  e^{i {\bf k}_F {\bf r}_{1 2}} G^R_c ({\bf r}_{1 2}, \mu) \Pi ({\bf r}_{1 2}, \mu) ,
\label{y2}
\eqa
Integrating each RG equation, we set the upper limit
 to the bare band width $D_0$ and
 the lower one to the respective energy scale $D = T_{ki}$,
 where $g_i$, $i=1,2$ is diverging.
 Thereby, we find  the two coupled equations for the Kondo temperatures
 $T_{K_1}$ and $T_{K_2}$
\begin{equation} \label{RG1}
\frac{1} {g^0_{1}} = -2 \ln \frac{T_{K1}}{D_0} + y g^0_{1} g^0_{2} \frac{D_0}{T_{K2}}  \ln ( \frac{ \sqrt{1+(T_{K1}/T_{K2})^2 }-1} { \sqrt{1+(T_{K1}/T_{K2})^2 }+1}         )
\end{equation}
\begin{equation} \label{RG2}
\frac{1} {g^0_{2}} = -2 \ln \frac{T_{K_2}}{D_0} + y g^0_{1} g^0_{2} \frac{D_0}{T_{K_1}}  \ln ( \frac{ \sqrt{1+(T_{K_2}/T_{K_1})^2 }-1} { \sqrt{1+(T_{K_2}/T_{K_1})^2 }+1}         )
\end{equation}
 Rescaling the Kondo temperature $T_{K_i}$
  with the bare Kondo temperatures as
\begin{equation}
x_i= \frac{T_{K_i}}{T^0_{K_{i}}},
\end{equation}
  where $0<x_i<1$, for  $i=1,2$,
  we can  rewrite   Eqs. (\ref{RG1},
  \ref{RG2}) as
\begin{equation}\label{RGx1}
2 \ln x_1 - \frac{\Tilde{y_1}}{x_0} \frac{1}{\alpha e} \frac {1}{x_2} \ln ( \frac{ \sqrt{1+(\frac{x_{1}}{x_{2}})^2 \frac{1}{x_0^2}   }-1} { \sqrt{1+(\frac{x_{1}}{x_{2}})^2 \frac{1}{x_0^2}   }+1}         )=0
\end{equation}
\begin{equation} \label{RGx2}
2 \ln x_2 - \Tilde{y_1} \frac{1}{\alpha e} \frac {1}{x_1} \ln ( \frac{ \sqrt{1+(\frac{x_{1}}{x_{2}})^2 x_0^2   }-1} { \sqrt{1+(\frac{x_{1}}{x_{2}})^2 x_0^2   }+1}         )=0
\end{equation}
 Here we introduced
$x_0$ as the ratio of the  bare Kondo temperatures
\begin{equation}
x_0= \frac{T^0_{K_{2}}}{T^0_{K_{1}}}.
\end{equation}
 The critical  ratio of  the bare Kondo temperature and
 the bare RKKY exchange  is given by
 $y_{ci}= T^0_{Ki}/(\alpha~ e g^0_1 g^0_2 D_0 )$, $i=1,2$
  where
$\alpha = ln(1+\sqrt{2}) $. In the following we use
the rescaled RKKY parameter $\Tilde{y_i} = y/y_{ci}$.

Now, we solve the coupled  Eqs.  (\ref{RGx1}) and (\ref{RGx2})
by the method of simplified Monte Carlo Research Algorithm, where we used the
GSL mt19937 algorithm  for generation of  random numbers.
For identical local density of states and exchange couplings, the   Kondo temperatures are the same,
$x_0 =1$ and  we recover the results of Ref. \onlinecite{Nejati2017},
where
 the Kondo temperature
 decreases  with RKKY coupling.
For  RKKY coupling exceeding the critical value,
$y > y_c,$ Eq. (\ref{yc}), there is no
 Kondo screening  anymore, and the two magnetic impurity spins are quenched by the RKKY coupling. At the critical value, $y_c$, Eq. (\ref{yc}),
 the Kondo temperature is
  reduced to $T_K(y_c) = e^{-1} T^0_K \approx 0.368 T^0_K$.

\begin{figure}
\includegraphics[scale=0.25]{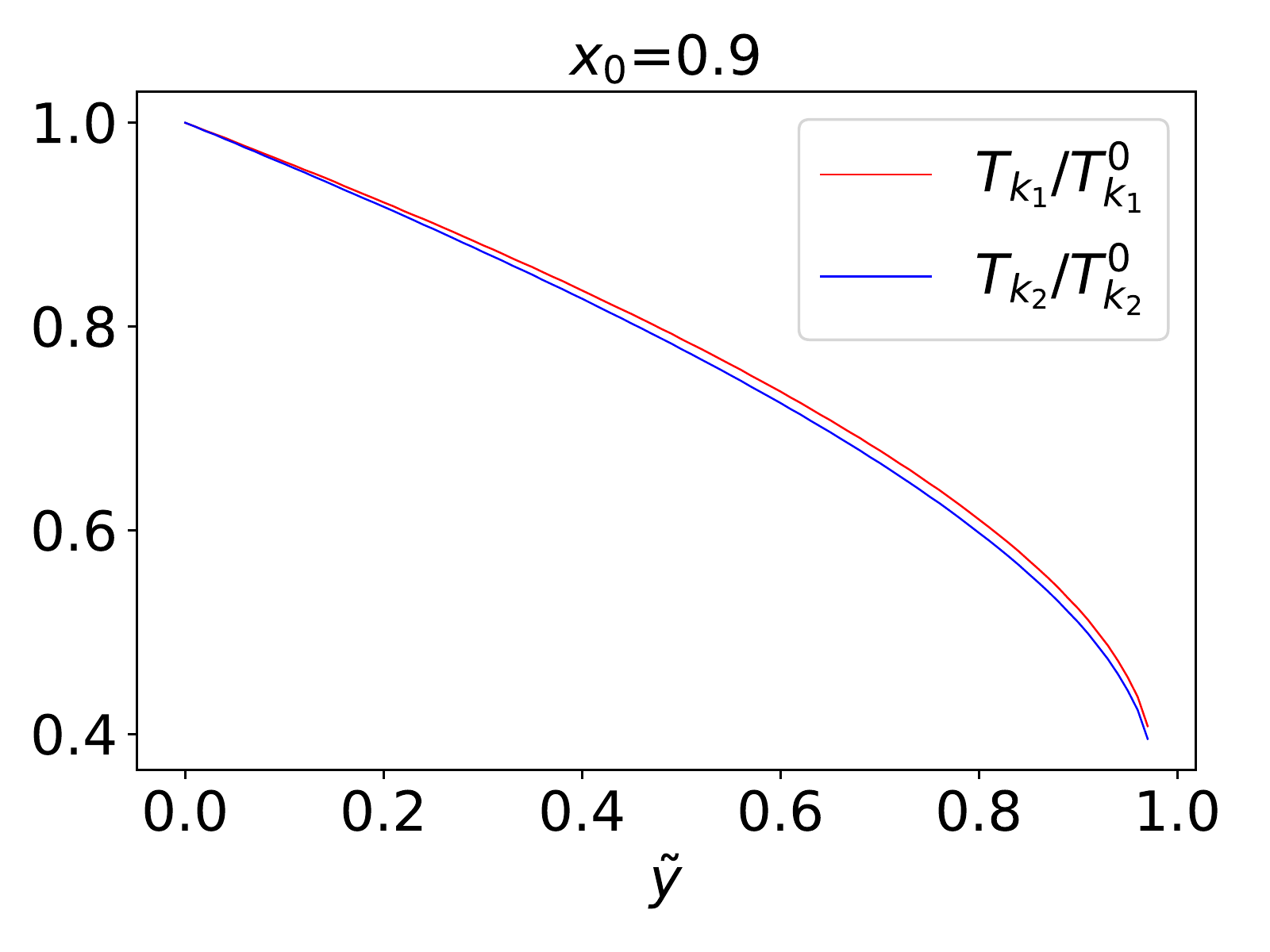}
\includegraphics[scale=0.25]{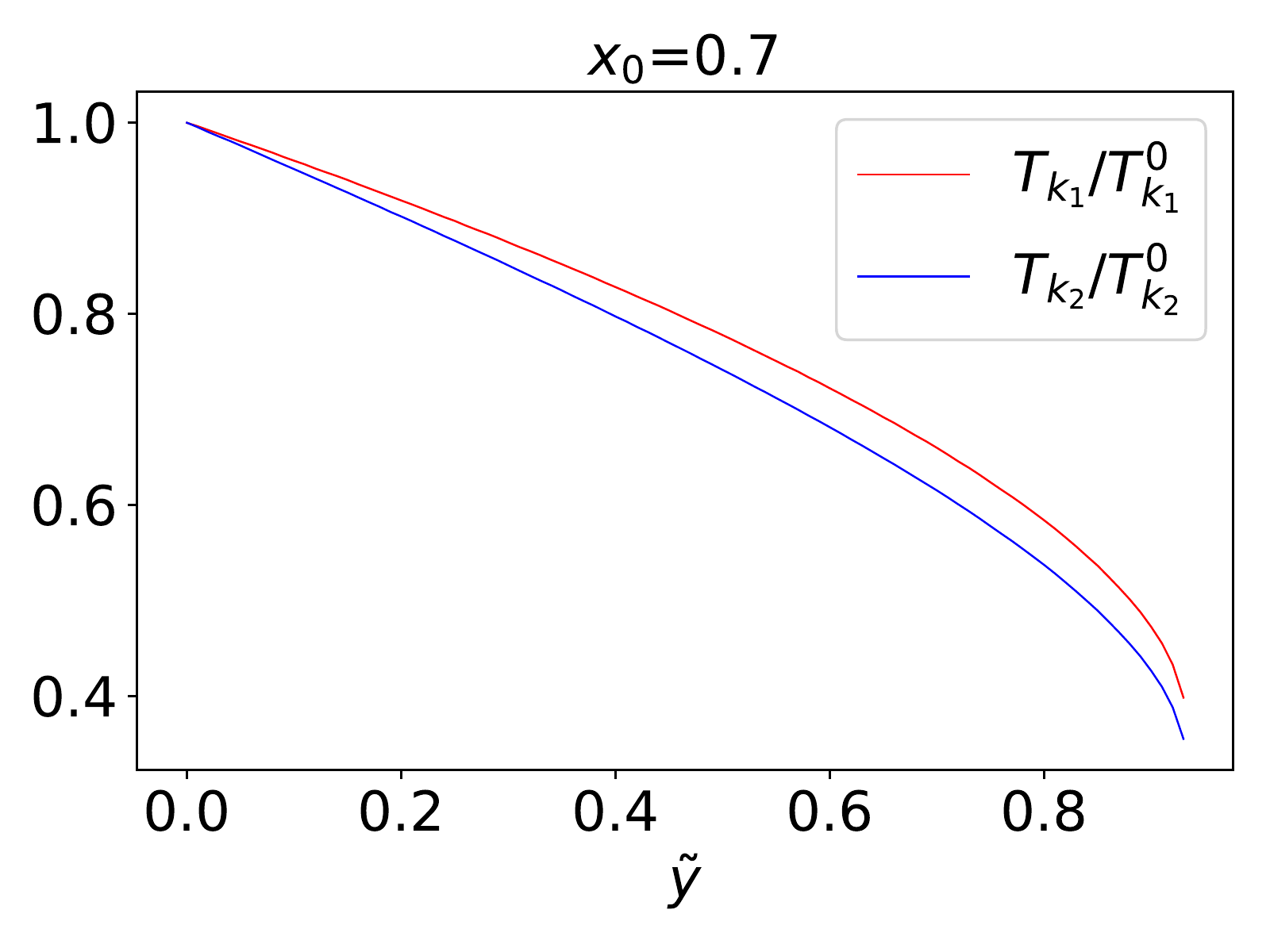}
\includegraphics[scale=0.25]{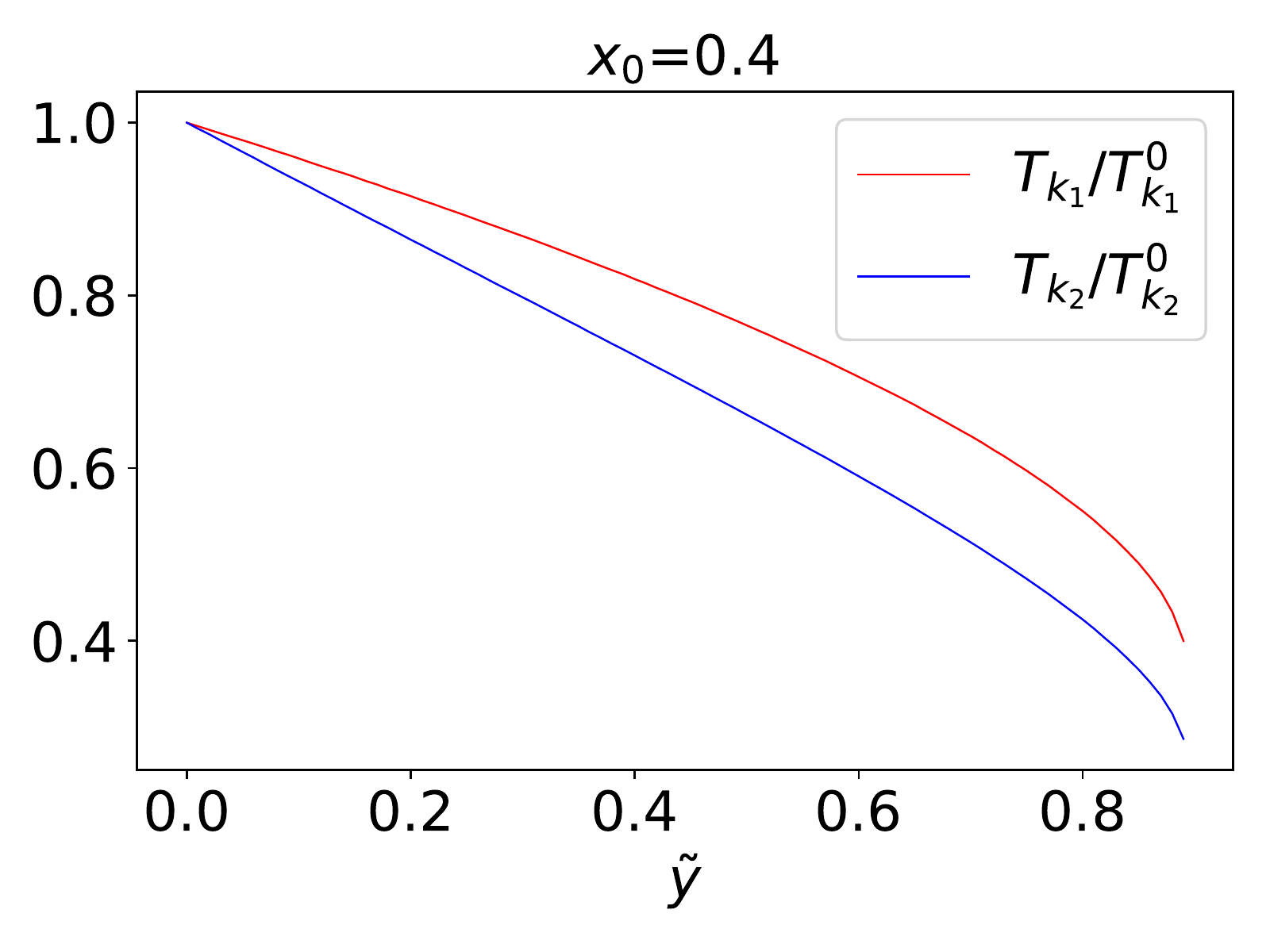}
\includegraphics[scale=0.25]{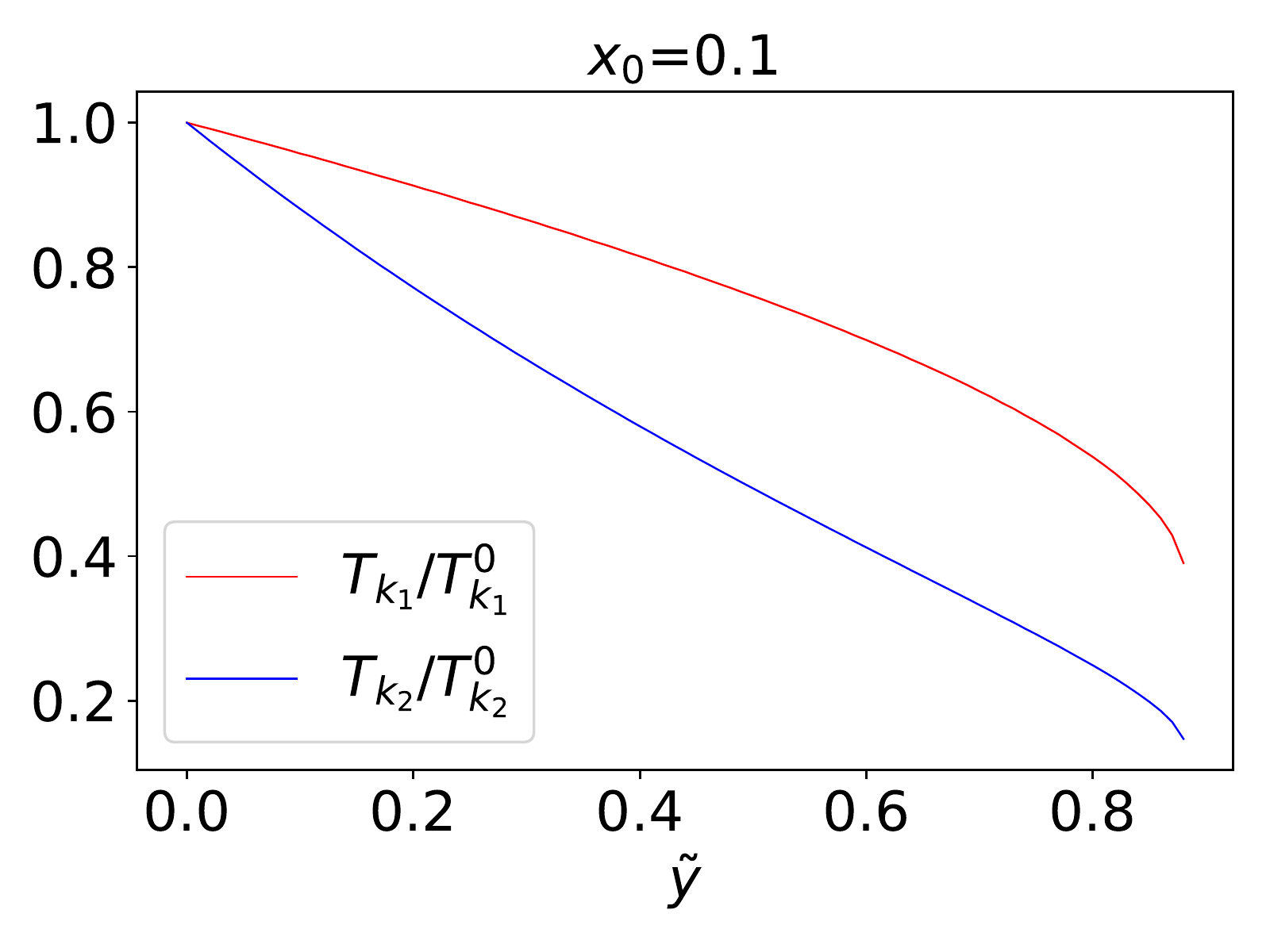}
  \caption{(Color online)  The Kondo temperatures  $T_{K1}$ and  $T_{K2}$
   of  two magnetic impurities
    relative to their bare values,
    as function of the dimensionless RKKY coupling parameter
    between them, $\tilde{y}$, relative to its critical value for the homogenous system, for  different bare Kondo temperature ratios $x_0= T^0_{K_{2}}/T^0_{K_{1}}  =0.9, 0.7, 0.4, 0.1$.  }
 \label{fig:rkkykondo}
\end{figure}

\begin{figure}
\includegraphics[scale=0.5]{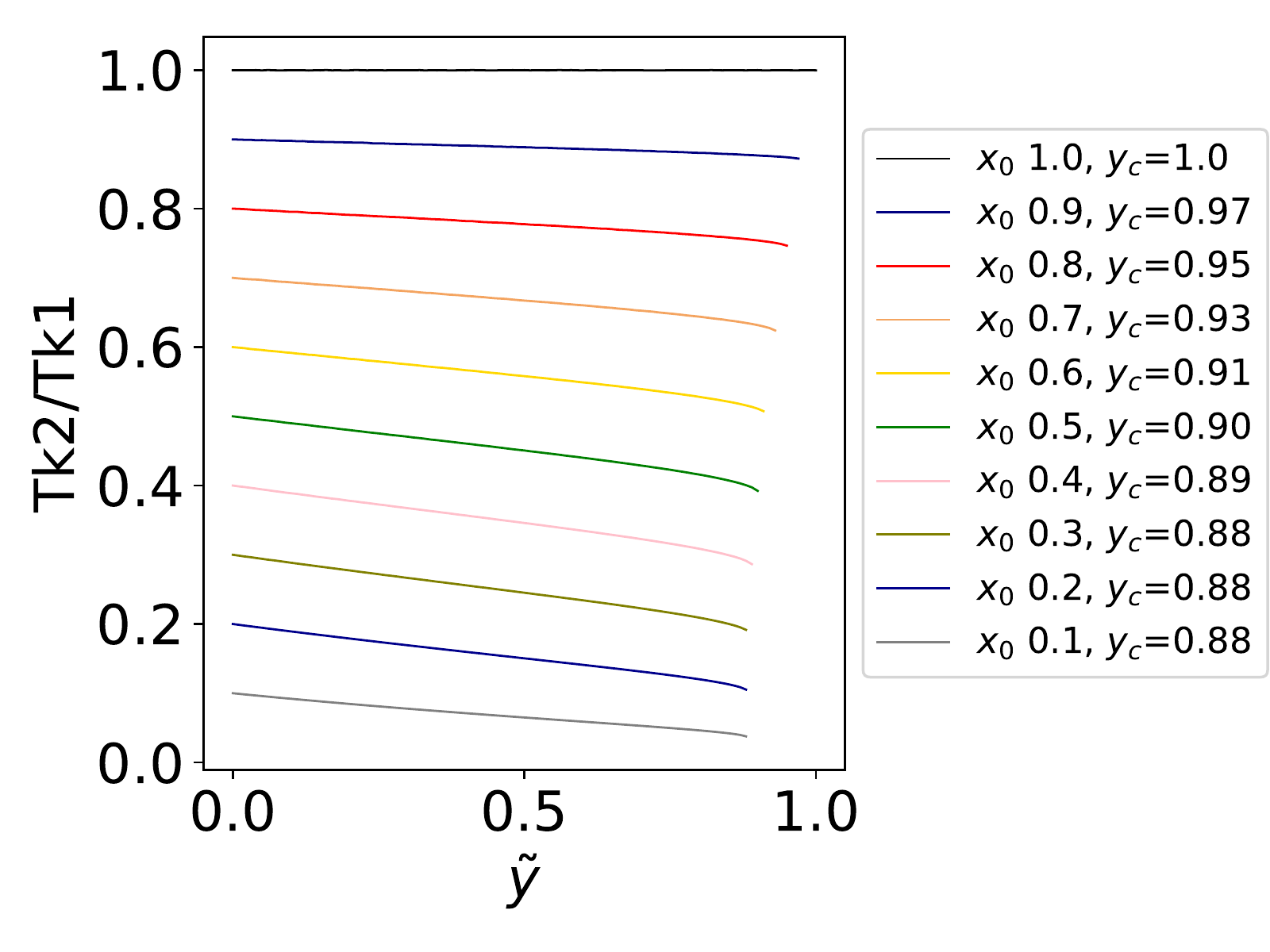}
  \caption{(Color online)  The ratio of
  Kondo temperature of  two magnetic impurities,
    $x=T_{K2}/T_{K1}$
    as function of the dimensionless RKKY coupling parameter $\tilde{y}$,
    relative to its critical value for the homogenous system,
    for different bare Kondo temperature ratios
    $x_0= T^0_{K_{2}}/T^0_{K_{1}}  =0.1, 0.4, 0.7, 0.9$. The curves end at a critical value $\tilde{y}_c (x_0)$, above which both Kondo impurities are quenched.  }
 \label{fig:rkkykondo2}
\end{figure}

\begin{figure}
\includegraphics[scale=0.25]{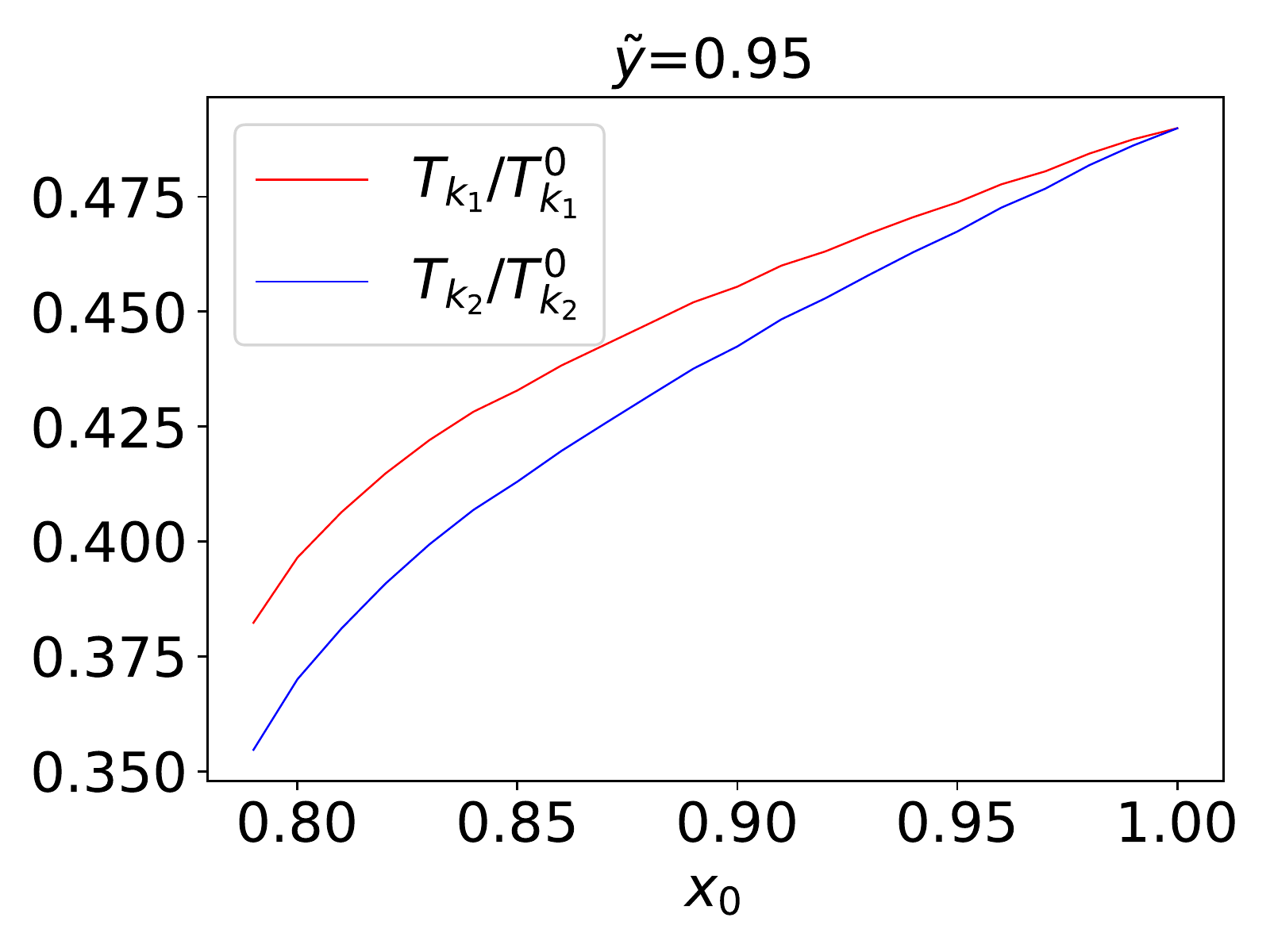}
\includegraphics[scale=0.25]{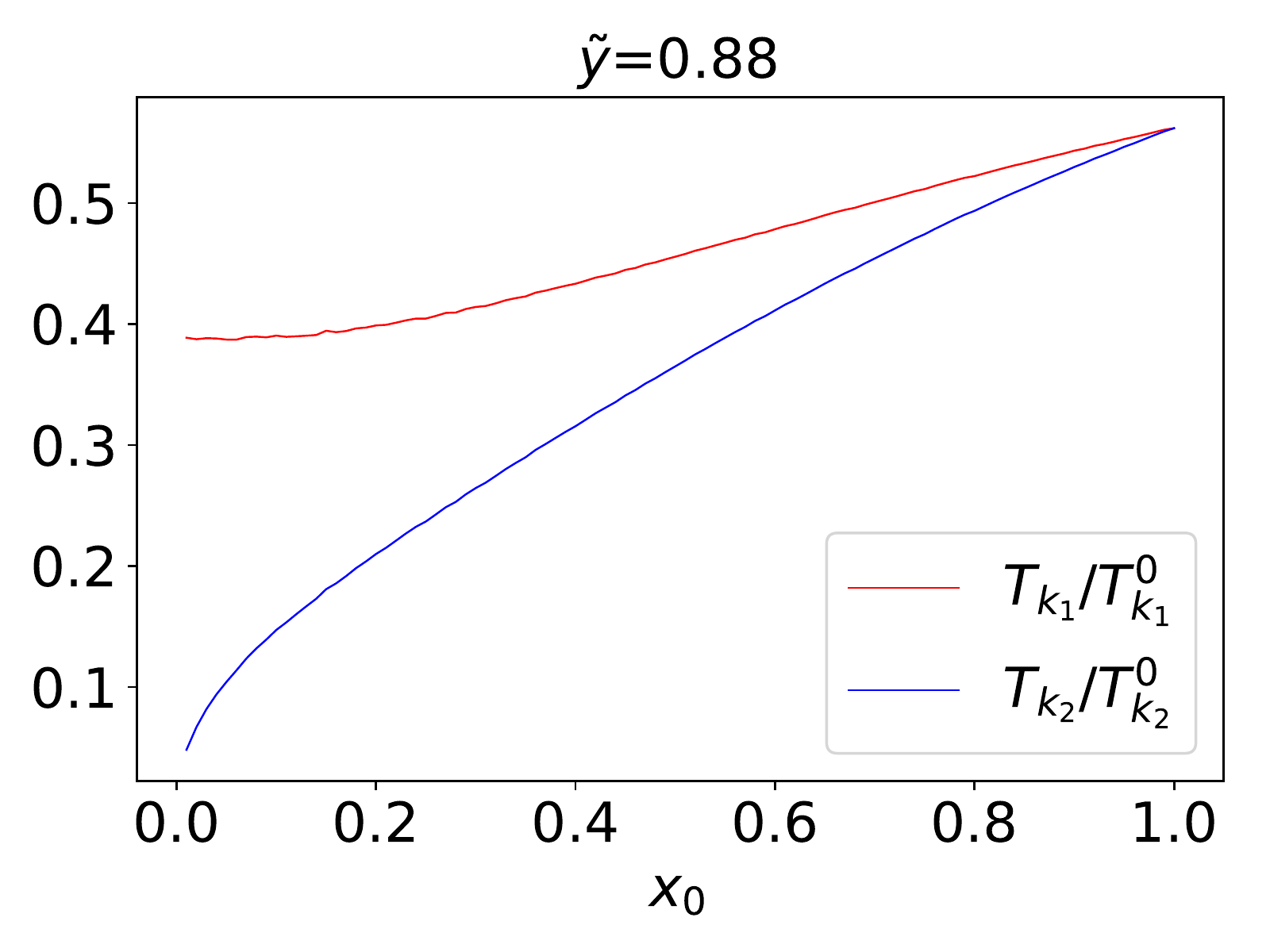}
\includegraphics[scale=0.25]{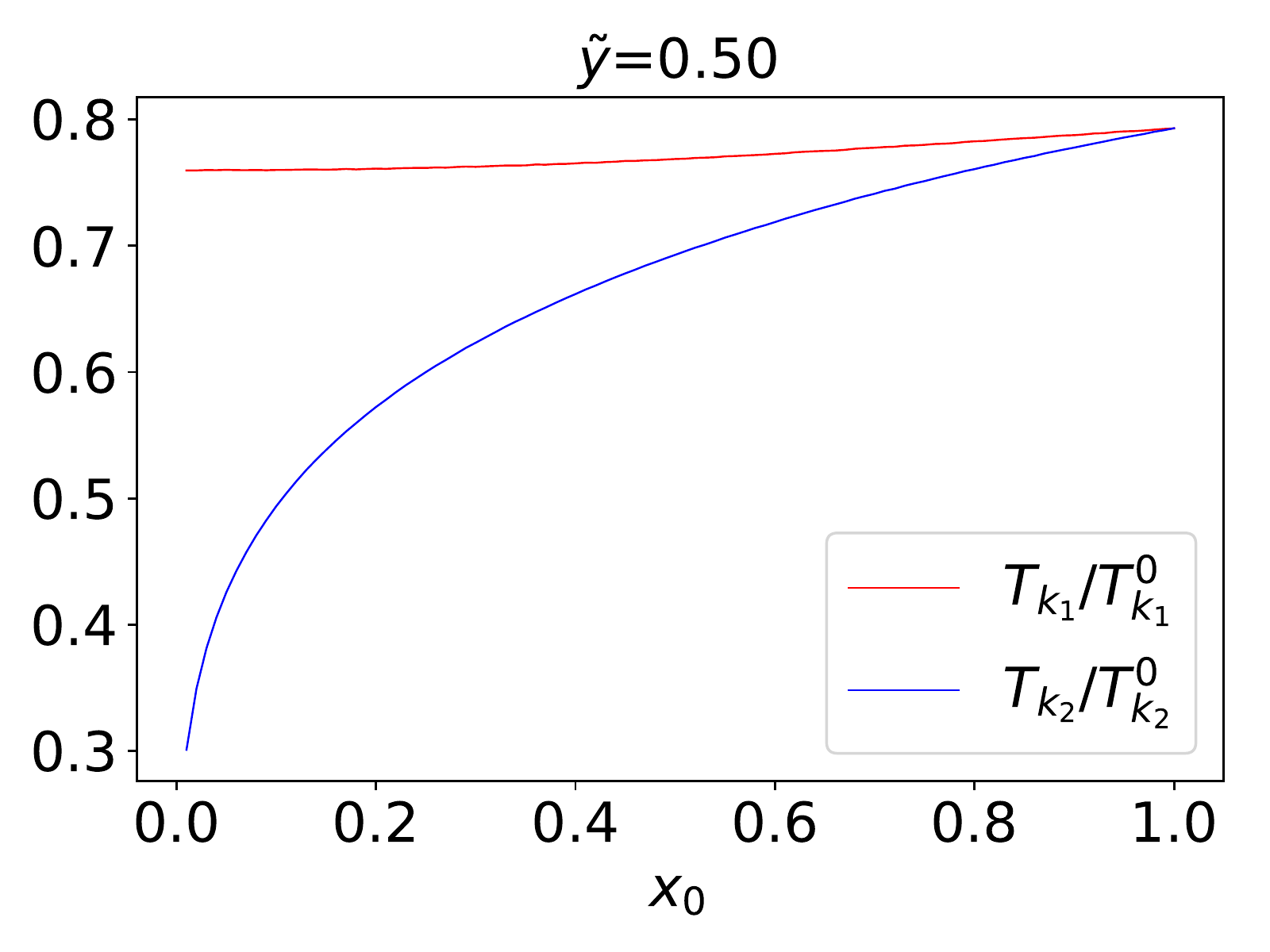}
\includegraphics[scale=0.25]{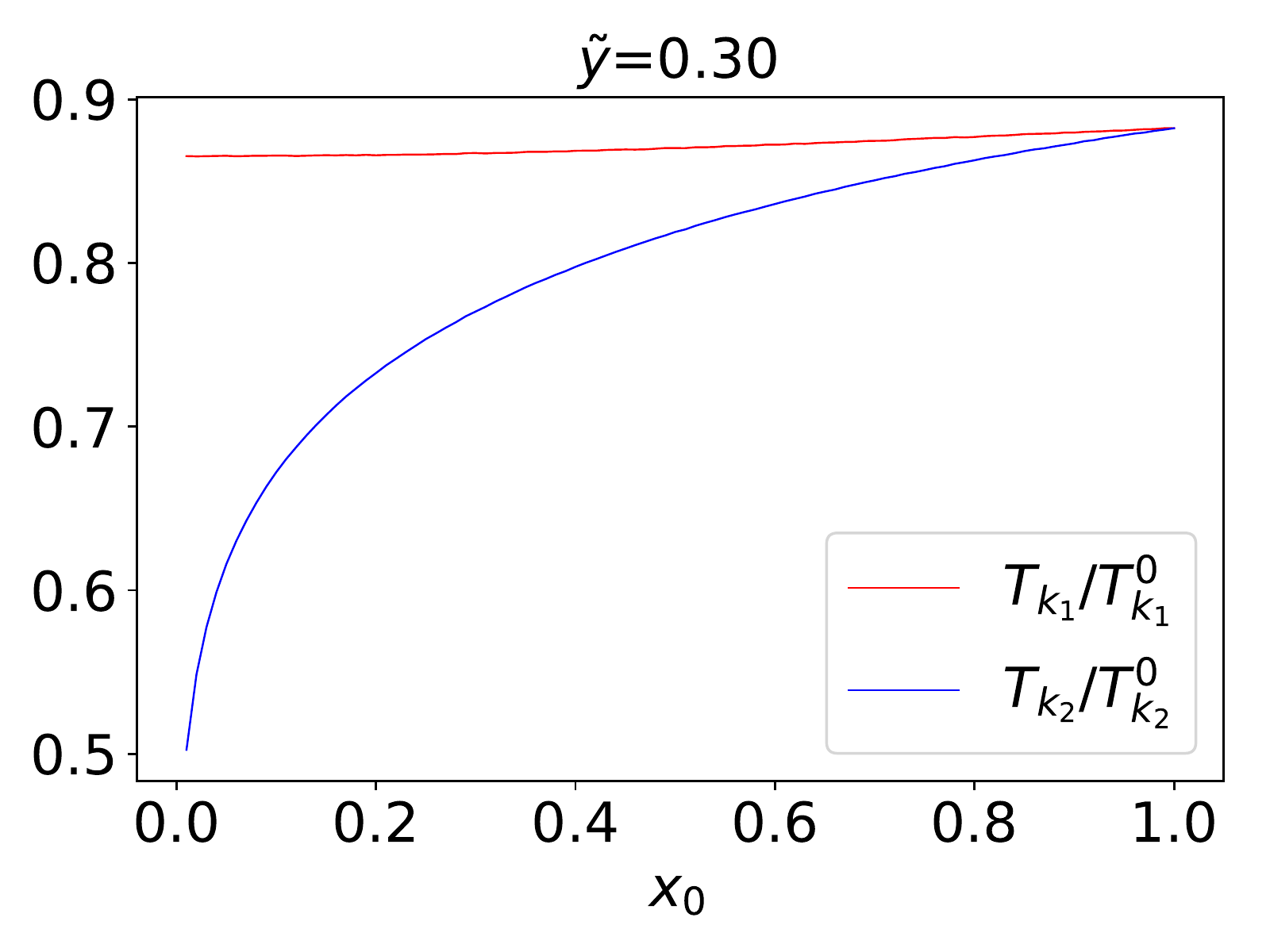}
\caption{(Color online) The
  Kondo temperatures of  two magnetic impurities,
    $T_{K1}$ and  $T_{K2}$ relative to their bare values
    as function of  bare Kondo temperature ratios
    $x_0$  for different  dimensionless RKKY coupling parameters $\tilde{y}=0.95,0.88,0.5,0.3$. }
 \label{fig:rkkykondo3}
\end{figure}

 Next,
let us consider  what happens  when the bare  coupling parameters
$g^0_{i}$, Eq. (\ref{g0}) and thereby the  bare Kondo temperatures  at the  two sites  are different.
  We take $x_0<1$, and solve the coupled  Eqs.  (\ref{RGx1}) and (\ref{RGx2})
  for  increasing values of the  RKKY coupling $y$.
  The numerical results show that the RKKY coupling reduces
 both Kondo temperatures, but   the initially smaller Kondo temperature becomes suppressed more strongly than the larger one, so that the ratio $x = T_{K_{2}}/T_{K_{1}}$  decreases
  further.
  This effect becomes more pronounced the smaller
  the ratio $x_0$ is, initially, as seen in
   Fig. \ref{fig:rkkykondo}, where the  Kondo temperatures  $T_{K1}$ and  $T_{K2}$
   of  two magnetic impurities
    relative to their bare values,
    as function of the dimensionless RKKY coupling parameter
    between them, $\tilde{y}$  is plotted for
     various values of $x_0$,
 in Fig. \ref{fig:rkkykondo2}, where the  ratio of
  Kondo temperatures
    $x$
is plotted     as function of  $\tilde{y}$ and
 in  Fig. \ref{fig:rkkykondo3}, where  $T_{K1}$ and  $T_{K2}$
   relative to their bare values, are plotted
    as function of
    $x_0$ for various
    dimensionless RKKY coupling parameter
    between them, $\tilde{y}$.
  Thus, we conclude that inhomogeneity is a relevant perturbation and the resulting inequality in the Kondo temperatures becomes enhanced  further by the RKKY coupling.

\begin{figure}
\includegraphics[scale=0.5]{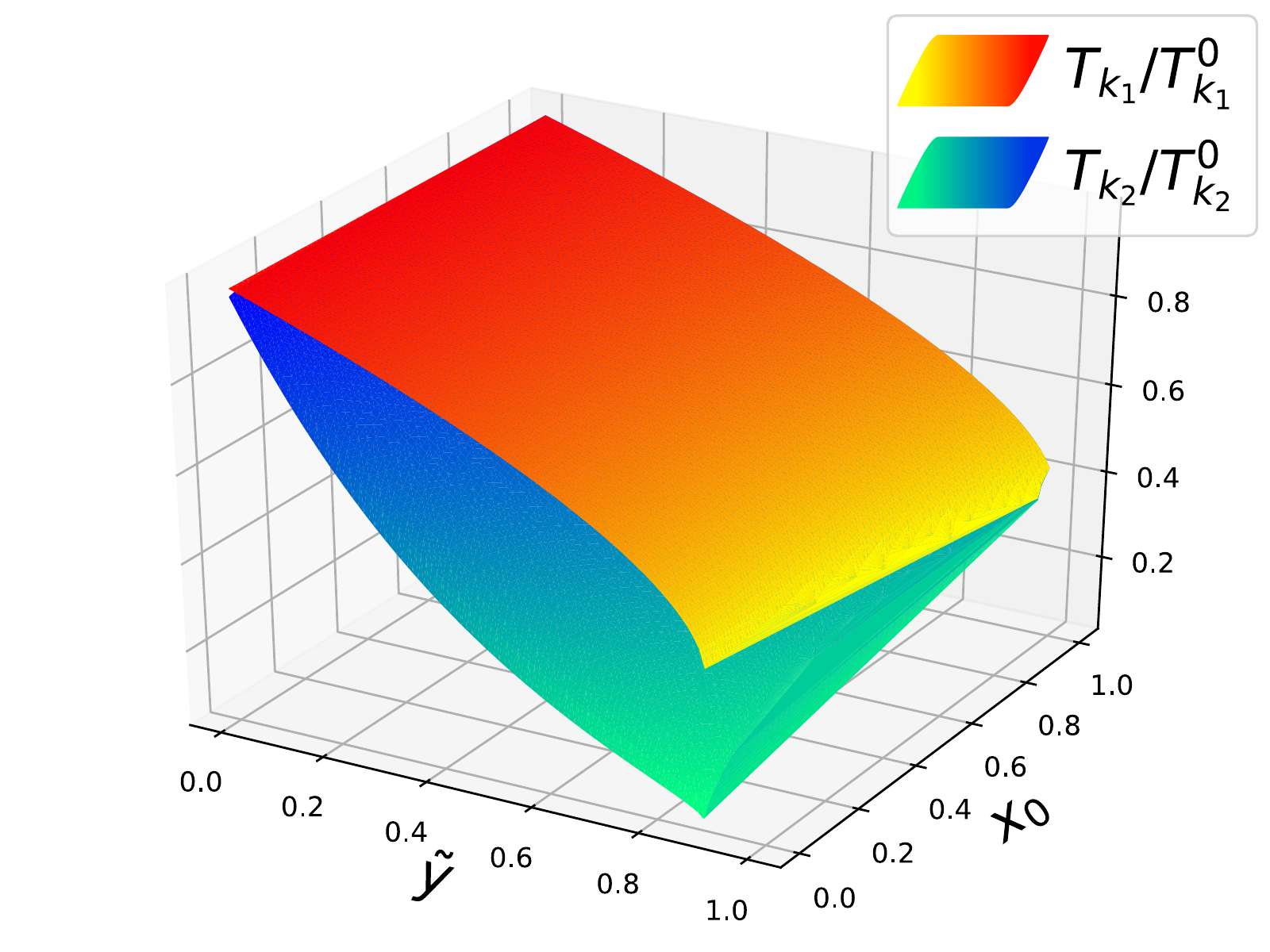}
\caption{(Color online)
Kondo temperatures of  two magnetic impurities,
    $T_{K1}$ and  $T_{K2}$ relative to their bare values
    as function of  bare Kondo temperature ratios
    $x_0$  and the dimensionless RKKY coupling parameters $\tilde{y}$.}
 \label{fig:rkkykondo4}
\end{figure}

\begin{figure}
\includegraphics[scale=0.5]{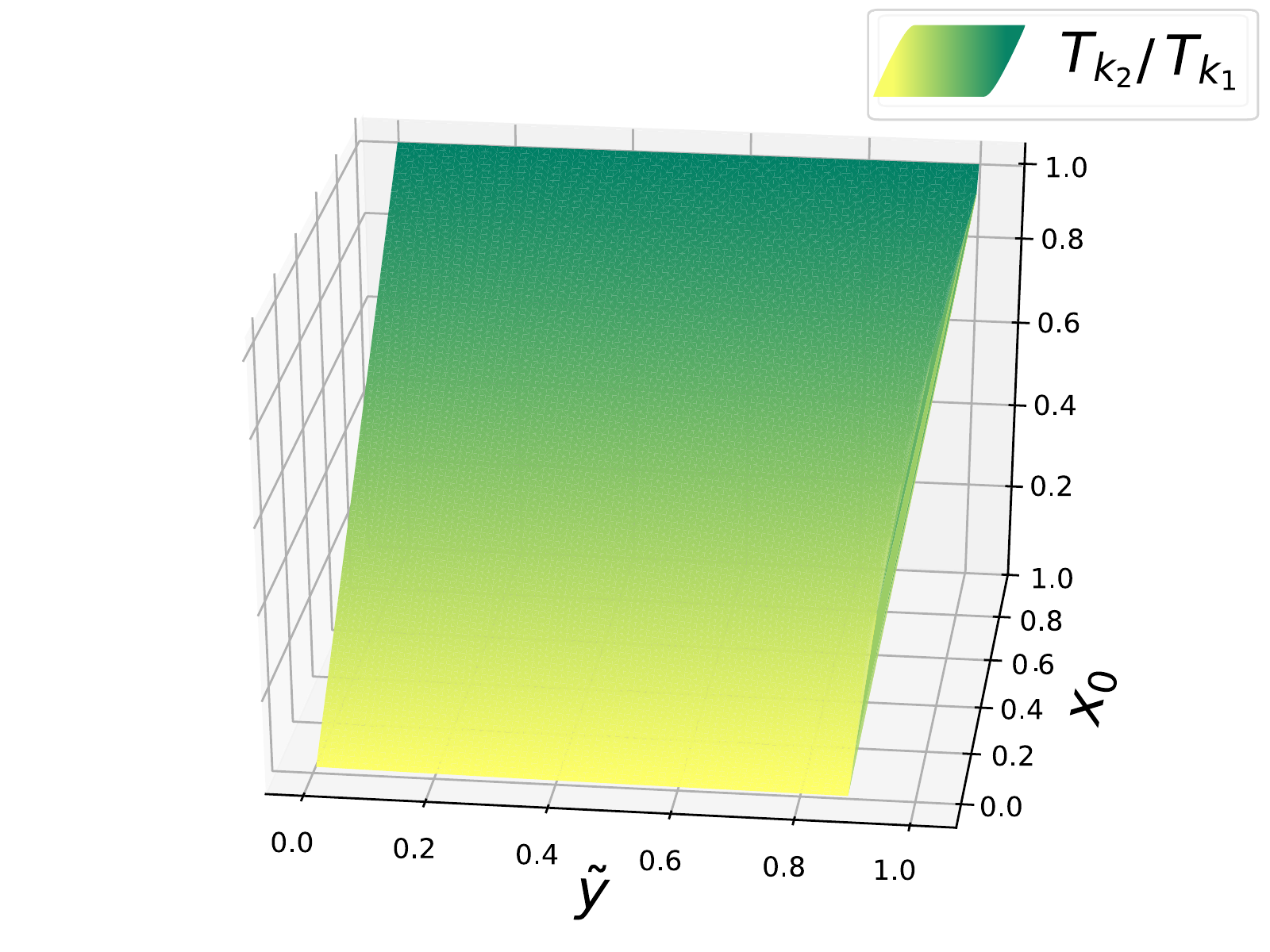}
\caption{(Color online)  Ratio of the Kondo temperatures of  two magnetic impurities,
    $T_{K1}/T_{K2}$
    as function of  bare Kondo temperature ratios
    $x_0$  and the dimensionless RKKY coupling parameters $\tilde{y}$. }
 \label{fig:rkkykondo5}
\end{figure}

Moreover,  the  quenching  of the Kondo screening by the RKKY coupling
 occurs already for smaller RKKY coupling, as seen in
 Fig.  \ref{fig:rkkykondo2},
  the stronger
  the inhomogeneity and the smaller the ratio of the bare Kondo temperature  $x_0$ is. For small $x_0$  the breakdown occurs
 at a critical value  which converges to
 $y_c(x_0 \ll 1)  = 0.88 y_c$  of the critical  RKKY coupling  $y_c$ in the homogeneous system, Eq. (\ref{yc}),
 as confirmed  in  table \ref{table1}.
   Thus, we find that
  inhomogeneity leaves the Kondo screening of the magnetic impurities   more easily quenchable by RKKY coupling.
 In   table  \ref{table1}  we  notice that,  while the larger of the two Kondo temperatures reaches at the critical coupling a value which is
  a little larger  than the value it would read in a homogenous system $T_K(y_c) = e^{-1} T^0_K \approx 0.368 T^0_K$, the smaller Kondo temperature reaches a much lower Kondo temperature than it could reach in the homogenous system.
   As  observed already above   in
   Fig. \ref{fig:rkkykondo} we also see that the smaller the initial ratio of Kondo temperatures $x_0$ is, the smaller the ratio of Kondo temperatures with coupling becomes, reaching for  $x_0=0.1$   a ratio $x_c = 0.038,$ just about one third
    of the initial ratio, confirming that the inequality between the Kondo temperatures becomes enhanced by the RKKY coupling.
    All that is seen in the three-dimensional plot
   Fig.  \ref{fig:rkkykondo4} where the  Kondo temperatures
    $T_{K1}$ and  $T_{K2}$ relative to their bare values are plotted
    as function of  bare Kondo temperature ratios
    $x_0$  and the dimensionless RKKY coupling parameters $\tilde{y}$,
     as well as in  Fig.  \ref{fig:rkkykondo5} where
      the ratio of the Kondo temperatures
    $x$  is plotted
    as function of
    $x_0$  and  $\tilde{y}$. In table \ref{table2} we list the values of
    $T_{K1}$ and  $T_{K2}$ and their ratio at the critical coupling
    for small $x_0.$ We see that the ratio  of the Kondo temperatures
    $x$ decays rapidly with $x_0.$

\begin{table}[]
\centering
\begin{tabular}{|c|c|c|c|c|}
\cline{1-5}
$x_0$ & $\tilde{y}_c$ & $T_{K1}(\tilde{y}_c)$/$T^{0}_{K1} $ & $T_{K2}(\tilde{y}_c)$/$T^{0}_{K2}$ & $T_{K2}(\tilde{y}_c)$/$T_{K1}(\tilde{y}_c)$
    \\\cline{1-5}
     1& 1  & 0.368  & 0.368  & 1 \\\cline{1-5}
     0.9& 0.97  &0.407   &0.395   & 0.872  \\\cline{1-5}
     0.8&0.95   &0.396   &0.370   & 0.746 \\\cline{1-5}
     0.7&0.93   &0.398   &0.355   & 0.634 \\\cline{1-5}
     0.6&0.91   &0.414   &0.350   & 0.517 \\\cline{1-5}
     0.5&0.90   &0.398   &0.312   & 0.392 \\\cline{1-5}
     0.4&0.89   &0.400   &0.286   &0.286 \\\cline{1-5}
     0.3&0.88   &0.414   &0.264   & 0.191 \\\cline{1-5}
     0.2&0.88   &0.399   &0.210   & 0.105 \\\cline{1-5}
     0.1&0.88   &0.391   &0.147   & 0.038 \\\cline{1-5}
\end{tabular}
\caption{The  Kondo temperatures at the respective critical coupling $\tilde{y}_c (x_0)$ and their ratio  as function of the ratio of bare couplings.}
\label{table1}
\end{table}

%
%

To confirm this anisotropic Kondo destruction, we consider the limit $x_{0} \rightarrow 0$, so that one of the two bare Kondo temperatures is vanishingly small. Then, Eq. \eqref{RGx1} is reduced to
\begin{equation} \label{t1}
2 \ln x_1 + \frac{2 \Tilde{y_1}}{x_1 \alpha e} + \mathcal{O}(x_0^2)
 =0
\end{equation}
and   Eq. \eqref{RGx2} to
\begin{equation} \label{t2}
2 \ln x_2 - \frac{4 \Tilde{y_1}}{x_1 \alpha e} \ln ( \frac{x_0 x_1}{2 x_2})+\mathcal{O}(x_0^2)
 =0.
\end{equation}
Solving Eq. \eqref{t1}, one finds $T_{K1}$ as a function of the effective RKKY interaction strength $\Tilde{y_1}$. Inserting the maximum value of $\Tilde{y_1}$ = $\alpha \sim 0.8813$, we obtain that $x_1$ remains to be finite and close to $1/e \sim 0.36$. Solving Eq. \eqref{t2}, we obtain $T_{K2}$ as a function of the ratio $x_{0}$ of bare Kondo temperatures. As a result, we have
%
\begin{equation} \label{t3}
 x_{2} = \left(  \frac{x_{0} x_{1}}{2} \right)^{\frac{\frac{2 \Tilde{y_1}}{x_1 \alpha e}}{1 + \frac{2 \Tilde{y_1}}{x_1 \alpha e}}},
\end{equation}
which  decays rapidly as $x_0 \rightarrow 0$, setting $x_1 = 1/e$ in accordance with the numerical solution tabled in table \ref{table2}.

\begin{table}[]
\centering
\begin{tabular}{|c|c|c|c|c|}
\cline{1-5}
 $\Tilde{y_1}$ & $x_0$ & $T_{K1}$/$T^{0}_{K1}$ & $T_{K2}$/$T^{0}_{K2}$ & $T_{K2}$/$T_{K1}$  \\\cline{1-5}
     0.881 & $10^{-3}$  & 0.372  & 0.0140  & $3.76 \times 10^{-2} x_0$  \\\cline{1-5}
     0.881& $10^{-4}$  &0.371   &0.00442  & $1.19 \times 10^{-2}  x_0$    \\\cline{1-5}
     0.881& $10^{-5}$  &0.366   &0.00133   & $3.63  \times 10^{-3} x_0$\\\cline{1-5}
     0.881& $10^{-6}$  &0.367   &0.000424   & $1.16 \times 10^{-3} x_0 $ \\\cline{1-5}
     0.881 & $10^{-7}$ &0.375   &0.000149 & $3.97 \times 10^{-4} x_0$  \\\cline{1-5}
     \end{tabular}
\caption{Numerical solutions
for Eqs. \eqref{RGx1} and \eqref{RGx2} in the $x_0 \rightarrow 0$ limit.}
\label{table2}
\end{table}

\section{The Kondo effect in the system of randomly distributed magnetic impurities}

Next, let us consider   the generalisation of this theoretical framework  to an electron system with a finite density of  randomly distributed magnetic impurities,  $n_M = N/Vol.$. As $N$ magnetic impurities
are placed at random positions
${\bf r}_i$, $i=1,...,N$, they are coupled by
random
local exchange couplings $J^0_i$ to the conduction electrons with   local density of states $\rho(E, {\bf r}_i )$.
 Thereby,  every magnetic moment placed at random  positions
  has a different Kondo temperature,
  yielding  a  distribution of Kondo temperatures \cite{mott,BhattFisher92,Lakner,KondoDisorder,Cornaglia2006,Mucciolo,Micklitz05,Kats,Slevin2019}.
 As  the RKKY coupling is  randomly distributed as well
 \cite{BhattFisher92,Lerner1993,HYLee2014}
 it remains an open problem to derive the quantum phase diagram of
 a disordered electron system with  finite density of magnetic moments $n_M$.
 Using the definition of the  RKKY couplings Eq. (\ref{y}),
we can now
generalize the self-consistent renormalisation group  equations
Eq. (\ref{rg2}) of the RKKY-coupled randomly distributed magnetic impurities.
It is important to note that  the   local density of states $\rho( {\bf r}_i,E )$ does depend on energy $E$, so that at each  RG scale $D$
the  renormalisation of the local  exchange coupling  $ J(\bm{r})$
 depends on the local density of states
 at energy $E=\mu \pm D$,
 $\rho( {\bf r}_i ,\mu \pm D)$,
  which may be different from the density of state at the chemical potential $\mu$, $\rho_{0 i} = \rho({\bf r}_i ,\mu)$.
Defining $g_i = J(\bm{r}_i) \rho_{0 i}$,  we thereby find for the renormalisation of exchange couplings $g_i$
\begin{widetext}
\bqa \label{rgnm1} && \frac{d g_i}{d \ln D} = -  g_i^{2}
\sum_{\alpha=\pm} \left(
\frac{\rho(\mu+ \alpha D,\bm{r}_i)}{\rho_{0 i}}
- \frac{4 J_i^0}{\pi \rho_{0i}}
  \sum_{j \neq i} J_j^0 {\rm Im} [e^{i {\bf k}_F {\bf r}_{ij} } \chi_c ({\bf r}_{ij},\mu+ \alpha D)  G^R_c ({\bf r}_{ij},\mu+ \alpha D) \chi_f ( {\bf r}_{j},\mu+ \alpha D ) ] \right)  . \eqa
\end{widetext}
 The first term on the right hand side corresponds to the 1-loop RG for the Kondo problem with energy dependent density of states
\cite{suhl,Zarand96,gapless}. In the second term,
$\chi_f ( {\bf r}_{j},E) $ is
the full f-spin susceptibility of the magnetic impurity positioned
 at ${\bf r}_{j}$.
$ G^R_c ({\bf r}_{ij},E)$ is the retarded   conduction electron propagator from position ${\bf r}_{i}$ to ${\bf r}_{j}$ with
${\bf r}_{ij} = {\bf r}_{i} - {\bf r}_{j}$.
$\chi_c ({\bf r}_{ij},E) $ is the conduction electron correlation function between positions ${\bf r}_{i}$ and ${\bf r}_{j}$.

 At moderate magnetic impurity densities $n_M$,
  one may approximate $\chi_f ( {\bf r}_{j},E) $  by the
  expression for a single Kondo impurity which is known from Bethe-Ansatz solution \cite{bethe}. This has been done in
   Ref. \cite{Nejati2017}, noting that only its
   real part contributes which is  then given by
  ${\rm Re} \chi_f ( {\bf r}_{j},\mu+D)  = W/(\pi T_{Kj} \sqrt{1+D^2/T_{Kj}^2}) $, where $W$ is the Wilson ratio and $T_{Kj}$ is the Kondo temperature of the magnetic impurity at position ${\bf r}_{j}$.

  In  Ref. \cite{Nejati2017} it has been furthermore assumed that
   all conduction electron properties, the local density of states,
   the propagator $ G^R_c ({\bf r}_{ij},E)$ and  the
   correlation function $\chi_c ({\bf r}_{ij},E) $ depend only
    weakly on energy, and therefore can be replaced by its value
     at the chemical potential $\mu$.  Let us therefore at first
      follow this assumption, although it is well known that
       the energy dependence can change the Kondo renormalisation \cite{gapless}
       and modify the distribution of Kondo temperatures in disordered systems substantially \cite{Mucciolo,Zhuravlev2007,Kats,Slevin2019}.
  Introducing furthermore the   continuum representation,
by
denoting $g(\bm{r}) = g_i$ and $\bm{r} =\bm{r}_i$,
 we can rewrite the RG equation  as
\begin{widetext}
\bqa && \frac{d g(\bm{r})}{d \ln D} = -  2 g(\bm{r})^{2} \left(1-
  g_{0} (\bm{r}) D_{0} \int d^{3} \bm{r}'  g_{0} (\bm{r'}) \frac{y(\bm{r}-\bm{r}')}{T_{K}(\bm{r}')} \frac{1}{\sqrt{1 + (D/T_{K}(\bm{r}'))^{2}}}\right). \eqa
\end{widetext} where we
introduced the function $y(\bm{r}-\bm{r}')$
defined by
\begin{widetext}
\bqa \label{yr}
y(\bm{r} -\bm{r}') = -
\frac{8 W}{\pi^2 \rho_0( \bm{r})  } {\rm Im} \sum_{j \neq i}
\frac{1}{ \rho_0( \bm{r_j})  } \delta ( {{\bf r}' - {\bf r}_j}) e^{i {\bf k}_F ({\bf r}-{\bf r}_j) } G^R_c ({\bf r}-{\bf r}_j, \mu) \chi_c ({\bf r} - {\bf r}_j, \mu).
\eqa
\end{widetext}
As a result, we obtain the self-consistent equation with RKKY interactions in the dilute limit of randomly distributed magnetic impurities,
\begin{widetext}
\bqa \label{tky} && - \frac{1}{g_{0}  (\bm{r}) } = 2 \ln \Big( \frac{T_{K}(\bm{r})}{D_{0}} \Big) - g_{0} (\bm{r}) D_{0} \int d^{3} \bm{r}' g_{0} (\bm{r}')  \frac{y(\bm{r}-\bm{r}')}{T_{K}(\bm{r}')} \ln \Big( \frac{\sqrt{1 + [T_{K}(\bm{r})/T_{K}(\bm{r}')]^{2}} - 1}{\sqrt{1 + [T_{K}(\bm{r})/T_{K}(\bm{r}')]^{2}} + 1} \Big) .
\eqa
\end{widetext}
Solving this integral equation, we can derive  the position dependent Kondo temperatures  for  a given configuration of RKKY interactions.
From the distribution of the
local couplings  $g_{0} (\bm{r})$ which originates from the random positions of doped magnetic impurities, the long range
function $y(\bm{r}-\bm{r}')$,  together with
the random  distribution of
 electronic properties like the local density of states, we can thus
 derive from Eq. (\ref{tky})
the  distribution function of Kondo temperatures $T_{K}$.
 We note that it has been found before that the random distribution of RKKY-coupling is mainly
 due to the distribution of local couplings $g_{0} (\bm{r})$ \cite{HYLee2014}, so that the distribution originates mainly from the
 local couplings  $g_{0} (\bm{r})$, while
 the function $y(\bm{r}-\bm{r}')$
  is not strongly modified by the disorder.

When the
  RKKY interaction is neglected
  it has been derived before that the Kondo temperature has a bimodal distribution with a  low $T_K$ peak and a peak close to the
   Kondo temperature of the clean systems \cite{KondoDisorder,Cornaglia2006,Mucciolo,Kats,Slevin2019}.
The low $T_K$-peak was found to become more pronounced   for stronger disorder and converges to a universal power law tail at the Anderson metal-insulator transition, where the power exponent depends only
 on the multifractality parameter $\alpha_0$ \cite{Kats,Slevin2019}.
   It remains to find out,  whether the relevance of inequalities found for two  magnetic impurities with RKKY-interaction above, where we found that the lower Kondo temperature is suppressed more strongly, results in a further  enhancement of the low Kondo temperature peak in its distribution and thereby of the low temperature magnetic susceptibility. This question can be resolved by the solution of
    Eq. (\ref{tky}), which we leave for further study.




\acknowledgments

This study was supported by the Ministry of Education, Science, and Technology (No. 2011-0030046) of the National Research Foundation of Korea (NRF). S.K. gratefully acknowledges support from DFG (Deutsche Forschungsgemeinschaft) KE-807/22-1. We gratefully acknowledge useful discussions with Keith Slevin.

\end{document}